\documentclass[12pt,preprint]{aastex}

\usepackage{pdflscape}
\usepackage{color}

\shorttitle{CMEs Deflection And Ambient Magnetic Field Configuration}
\shortauthors{Wang et al.}

\begin{document}

\title{On Deflection of Solar Coronal Mass Ejections by the Ambient Coronal Magnetic Field Configuration}
\author{
Jingjing Wang\altaffilmark{1},
J. Todd Hoeksema\altaffilmark{2},
Siqing Liu\altaffilmark{1,3}}
\altaffiltext{1}{National Space Science Center, Chinese Academy of Sciences, Beijing, China.}
\altaffiltext{2}{W. W. Hansen Experimental Physics Laboratory, Stanford University, Stanford, CA 94305-4085.}
\altaffiltext{3}{University of Chinese Academy of Sciences, Beijing, China.}

\begin{abstract}

Solar Coronal Mass Ejections (CMEs) are sometimes deflected during their propagation. This deflection may be the consequence of interaction between a CME and a coronal hole or the solar wind. We analyze 44 halo-CMEs whose deflection angle exceeds 90$^\circ$. The coronal magnetic field configuration is computed from daily synoptic maps of magnetic field from SOHO/MDI and SDO/HMI using a Potential Field Source Surface (PFSS) model. By comparing the ambient magnetic field configuration and the measured position angles (MPA) of the CMEs, we conclude that the deflection of 80\% of the CMEs (35 of 44) are consistent with the ambient magnetic field configuration, agreeing with previous studies. Of these 35, 71\% are deflected toward the heliospheric current sheet (HCS), and 29\% toward a pseudo-streamer (PS), the boundary between the same-polarity magnetic field regions. This implies that the ambient coronal magnetic field configuration plays an important and major role in the deflection of CMEs, and that the HCS configuration is more important than PS. If we exclude 13 CMEs having much higher uncertainty from the sample, the agreement between the deflection of CMEs and the ambient field configuration increases substantially, reaching 94\% in the new sample of 31 CMEs.

\end{abstract}


\section{Introduction}
\label{s:int}

Coronal Mass Ejections \citep[CMEs;][]{gosling1974mass,gosling1990coronal,tsurutani1997interplanetary,schwenn2000heliospheric,richardson2001sources} are enormous eruptions of plasma ejected from the Sun into interplanetary space and are often associated with flares and eruptions of filaments. CMEs propagating near the Sun-Earth line frequently appear as haloes or partial haloes in plane-of-the-sky coronagraph images. Halo CMEs often arrive at Earth and can cause intense geospace environmental disturbances, which may affect infrastructure systems and technologies in space and on Earth \citep[e.g.][]{gonzalez1999interplanetary,horne2013space,macalester2014extreme,2015ApJ...809L..34L,riley2018extreme}.

Sometimes CMEs are observed to be deflected during their propagation \citep[e.g.][]{2009JGRA..114.0A22G,gui2011quantitative,wang2011statistical,lugaz2012deflection, wang2015role, kay2017deflection,yang2018filament}. The deflection can happen near the Sun or in interplanetary space. For deflection in interplanetary space, it is suggested that interaction between a CME and the large-scale background solar wind plays a role \citep[e.g.][]{2004SoPh..222..329W,lugaz2012deflection}. For the near-Sun deflection, \citet{2009JGRA..114.0A22G} suggest that nearby coronal holes (CHs) may cause CMEs to generally move away from open magnetic field regions. \citet{wang2011statistical} statistically analyzed 1078 CMEs from the Coordinated Data Analysis Workshop (CDAW) CME catalog during 1997-1998 and found that most CMEs are deflected toward the equator in the plane of the sky near solar minimum \citep[see also][]{gui2011quantitative,shen2011kinematic,zuccarello2011role,lynch2013sympathetic,2013SoPh..284...59M}. \citet{makela2013coronal} defined a CH influence parameter that depends on the CH area, average magnetic field strength, and distance from the CME location to describe the influence of CME deflection. Limitations of these studies are (1) only CHs in the Earth-side surface are taken into account while the far-side CHs are neglected; and (2) propagation of CMEs is three dimensional in nature, but is analyzed in two dimensions there.

In this paper, we carry out a comprehensive investigation of large CME deflections near the solar surface by taking into account the global magnetic field configuration and the three-dimensional nature of CME propagation. We calculate the global magnetic field from daily synoptic maps of magnetic field using a Potential Field Source Surface (PFSS) model \citep{schatten1969model,smith2001heliospheric} to determine the ambient magnetic field structures of CMEs. To determine CME deflection angle, we assume that CMEs initially erupt radially from their source location and appear at the same position angle (PA) in the plane of the sky when observed by the SOHO/LASCO \citep{brueckner1995large,domingo1995soho} coronagraph. If a CME is deflected near the Sun, the CME may appear at a different measured position angle (MPA) in the SOHO/LASCO coronagraph. The deflection angle is defined as the separation angle between the radial PA and MPA.

The paper is organized as follows. We describe data and methodology in Section 2. Results are presented in Section 3. In Section 4, we summarize and discuss the results.

\section{Data and methodology}
\label{s:method}

\subsection{Deflection angle and CME selection}
\label{s:twod}

We assume that a CME will initially erupt radially from the identified source (e.g. flare or filament). The original CME should propagate along a line drawn radially outward from the source location. Then we project the line into the two-dimensional plane of the coronagraph image. The projected radial PA of the line is determined by the location of the source and the observer. The PA is measured counter-clockwise from solar north. If a CME is deflected near the Sun, the projected measured position angle (MPA) in SOHO/LASCO coronagraph observations may be different from the radial PA.

We select CMEs from the halo-CME catalog in the {\it Coordinated Data Analysis Workshop} \citep[CDAW;][]{gopalswamy2009soho} database. The catalog contains the source and projected CME information. The properties of the deflected CMEs listed in this catalog include the source location, MPA, and the projected linear speed along the MPA. We define the deflection angle in two dimensions as the separation angle between the MPA and the radial PA. The deflection angle is positive if the MPA is in the counter-clockwise direction from the radial PA, and negative otherwise. It should be noted that, the deflection angle in the-plane-of-the-sky is a combination of the three-dimensional radial direction and the projection effect from the observed point.

The selection criteria are (1) CMEs are halo, (2) the deflection angle exceeds 90$^\circ$, and (3) CMEs occur from Jan, 1997 to Dec, 2018. Ultimately we identify 44 CMEs and list them in Table\, \ref{events}. The CME at 10:04 UTC, Oct 15, 1998 is not included due to a data gap in SOHO/MDI. The CMEs in this table are  deflected significantly in two dimensions. If a CME erupts from a source location and appears at a MPA that is at least 90$^\circ$ away, the deflection angle in three dimensions must at least exceed the angular distance from the CME location to the center of the solar disk. We take that angular distance as minimum deflection angle in three dimensions in Table\, \ref{events}.

For example in Table\, \ref{events}, the first CME erupted from S18E20 at 16:58 UTC, April 29, 1998, and so had a radial PA of 134$^\circ$; However, the MPA was 336$^\circ$. Therefore, the deflection angle in two dimensions is -158$^\circ$. The angular distance from the CME location to the center of the solar disk is 27$^\circ$. Thus, the real deflection angle in three dimensions must be at least 27$^\circ$.

\subsection{Large-scale coronal magnetic field structures}
\label{s:footp}

We employ a PFSS model \citep{altschuler1969magnetic,schatten1969model} to compute the global magnetic field configuration using daily synoptic maps of magnetic field from SOHO/MDI \citep{scherrer1995solar} and SDO/HMI \citep{schou2010polarization}. The polar field is corrected using the method suggested by \citet{sun2011new}. In the PFSS model it is assumed that the magnetic field is potential everywhere between the photosphere and a spherical source surface. The modeled field matches the radial component on the photosphere and is forced to become purely radial on the source surface. The source surface is defined as 2.5 solar radii (denoted by Rss) here. The order of harmonic coefficients computed is 120. 

\begin{figure}[t]
\centering
\includegraphics[width=1.\textwidth]{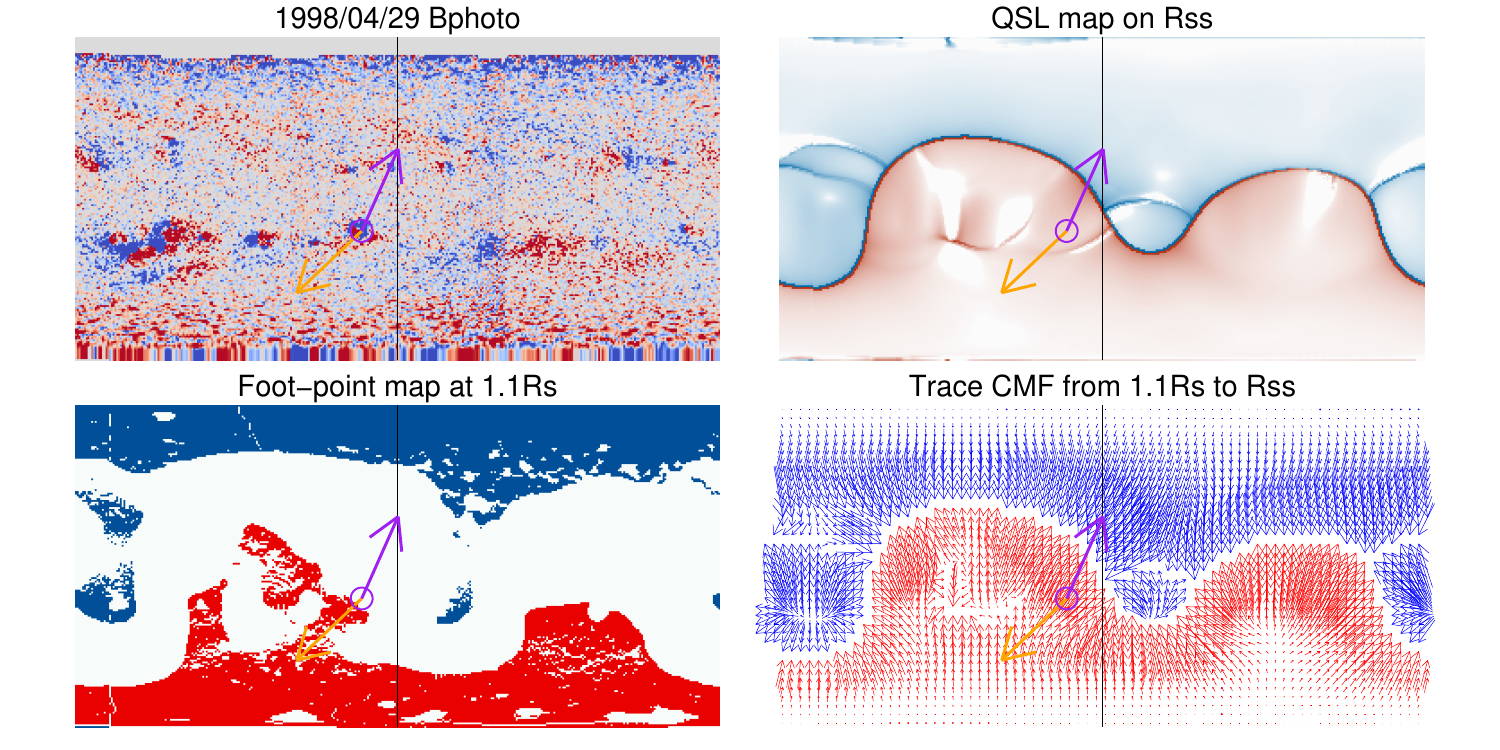}
\caption
{
\small Illustration of the large-scale coronal magnetic field (CMF) structure derived by PFSS from daily SOHO/MDI radial magnetic field synoptic map (upper left panel) in Carrington coordinates. The upper right panel shows the QSL map at the source surface (2.5 Rs). The lower left panel is the foot-point map at 1.1 solar radii. The lower right panel is the two-dimensional projected vector tracing the CMF from 1.1 Rs to the source surface at 2.5 Rs. The black vertical line in the middle refers to the central meridian on April 29, 1998. The blue and red color refer to positive and negative polarity, respectively. The purple circle indicates the source location of the CME that erupted at 16:58 UTC, April 29, 1998. The orange and purple arrows refer to the radial PA and the MPA of the CME determined using SOHO/LASCO coronagraph observations, respectively.
}
\label{illu}
\end{figure}

Figure\, \ref{illu} displays the large-scale magnetic field structure for the event occurring at 16:58 UT, April 29, 1998.  The top left panel is MDI daily synoptic map of the radial magnetic field on April 29, 1998. The blue and red colors refer to positive and negative fields. The bottom left shows footpoints of open flux traced from the source surface to 1.1 solar radii (Rs). The black vertical line refers to the central meridian at the start time of the CME. We further characterize the magnetic field configuration by computing the Quasi-Separatrix Layers \citep[QSL;][]{titov20081997,titov2011magnetic} from the PFSS field. The top right panel shows the QSL map at the source surface. The line separating red and blue areas is the base of the heliospheric current sheet \citep[HCS; e.g.][]{hoeksema1983structure}, the boundary between opposite polarities of magnetic field at the source surface. The separation lines within the red or blue regions indicate pseudo-streamers \citep[PS;][]{wang2007coronal,wang2019observations}, which are boundaries between same-polarity magnetic field regions.

We trace each open field line from 1.1 Rs to the source surface, and then draw a vector connecting the starting and ending points, as shown in the bottom right panel in Figure\, \ref{illu}. Coronal field lines that are close to radial appear as points. Strongly curved loops appear as longer vectors. The boundaries of HCS and PS are consistent with the QSL map on the source surface. The vectors show how the magnetic field lines evolve with height under the boundaries. This vector map is used to compare the ambient magnetic field configuration with the CME's MPA. The location of the 1998 April 29 16:58 CME is denoted with a purple circle. The orange and purple arrows refer to the radial PA and the MPA of the CME, respectively. This CME originates at the edge of a coronal hole (lower left) and emerges through a volume of the corona where the vectors are pointing toward the HCS (upper right). The purple arrow shows that the CME has been deflected toward the HCS in a direction consistent with the orientation of the large-scale ambient coronal field.

In order to compare the CME's MPA with the direction of the ambient coronal magnetic field (CMF) in a quantitative way, we define the direction of the ambient CMF by averaging the open magnetic field directions from the nine points within 1-degree of the CME location at 1.1 Rs. The separation angle between the MPA of the CME and the ambient CMF is 19$^\circ$. If the separation angle is less than 90$^\circ$, it is considered that the MPA of CME aligns with the direction of the ambient CMF. Therefore, for this event, the MPA of CME is consistent with the direction of the ambient CMF. This CME is deemed to be deflected in the direction of the ambient magnetic field configuration toward the HCS.

\section{Results}
\label{s:res}

For each event in the sample of 44 CMEs we repeat the analysis described in Section\, \ref{s:footp}. The tracing CMF maps, similar to the bottom right panel in Figure\, \ref{illu}, are shown in Figures\, \ref{case1}, \,\ref{case2}, and \,\ref{case3}. The final results are given in Table\, \ref{events}. The first column is the event number. Columns 2 to 6 are data from CDAW, listing properties of each CME. Column 7 is the radial PA. The minimum and projected deflection angles are in Columns 8 and 9. The minimum deflection angle is defined as the angular distance between the CME location and the center of the solar disk. The projected deflection angle is the deflection angle measured in the plane of the sky. The separation angle, defined as the angle between CME MPA and the direction of the ambient CMF, is given in Column 10. We hypothesize that the ambient coronal field deflects the CME, so that the MPA of the CME should be consistent with the direction of the ambient CMF. For the deflected CMEs that are consistent with the hypothesis, Column 11 indicates whether the CME is deflected toward the HCS or a PS. If this column is blank, it means that our hypothesis fails in this event.

In this sample of 44 CMEs, 35 CMEs support our hypothesis that the ambient coronal field deflects the CME. It is about 80\%. Among those, 25 CMEs (71\%) are deflected toward the HCS and 10 CMEs (29\%) toward a PS. This implies that the ambient coronal magnetic field configuration plays an important and major role in the deflection of the CMEs, and the HCS configuration is more important than PS.

There are nine cases that are opposite to our hypothesis. Among them, three cases (CME8, CME36, and CME41) are associated with eruption of filaments. Given the fact that there are only five filament-related cases in the entire sample, the disagreement rate is 60\%. One possible reason for this high percentage of disagreement may be that determining the location of filament eruption has greater uncertainty. Some of the locations may be mis-determined. Alternatively, filament-associated eruptions might be less affected by the overlying CMF. 

There are four cases (CME 22, CME27, CME31, and CME 41; including one filament-related case) in the disagreement group for which the events actually took place directly under the HCS or PS. In this situation, our method to derive the average direction of CMF has higher uncertainty. The method chooses the nine points nearest to the eruption location for averaging. Because the locations are under the HCS (or PS), the directions of the nine points are inconsistent, thus the final averaged direction has high uncertainty. 

There are six cases (CME8, CME13, CME18, CME34, CME35, and CME38; including one of the filament-related cases) whose minimum deflection angle as listed in Table\, \ref{events} is less than 10$^\circ$. Even a small deflection in disk-center CME can have a large apparent deflection, thus the projected deflection angle has high uncertainty.

There are five filament-related cases, four cases (including one of the filament-related cases) under HCS/PS, and six cases (including one of the filament-related cases) with minimum deflection angle less than 10$^\circ$. If we exclude these 13 higher uncertainty cases from the sample, we find that 29 of 31 cases are consistent with the hypothesis that the ambient magnetic field structure determines the deflection of CMEs. It is 94\% in agreement, strongly supporting our scenario.
      
\section{Conclusions and Discussion}
\label{s:con}

This investigation considers the CME deflection near the Sun. Using catalog information giving the CME location and the observed SOHO/LASCO coronagraph direction, we determine the measured position angle (MPA) and deflection angle of CMEs relative to radial propagation. We identify 44 halo-CMEs whose deflection angle in two dimensions exceeds 90$^\circ$. In order to explore the relationship between CME deflection and the ambient magnetic field structure, we employ a PFSS model to compute the coronal magnetic field from daily SOHO/MDI and SDO/HMI synoptic maps.

By comparing the ambient magnetic field configuration and the CMEs' MPA, we find that the deflections of 80\% of the CMEs are consistent with the ambient magnetic field configuration, agreeing with previous studies \citep{2009JGRA..114.0A22G,gui2011quantitative,zuccarello2011role,lynch2013sympathetic,2013SoPh..284...59M}. Of the consistently deflected CMEs, 71\% approach the heliospheric current sheet (HCS), the boundary between the magnetic field polarities, and 29\% move toward a pseudo-streamer (PS), the boundary between the same-polarity magnetic field regions. 80\% of the CMEs are deflected toward a large-scale coronal magnetic field. This indicates the coronal magnetic field configuration plays an important role in the deflection of CMEs.

\begin{figure}[!h]
\centering
\includegraphics[width=1.\textwidth]{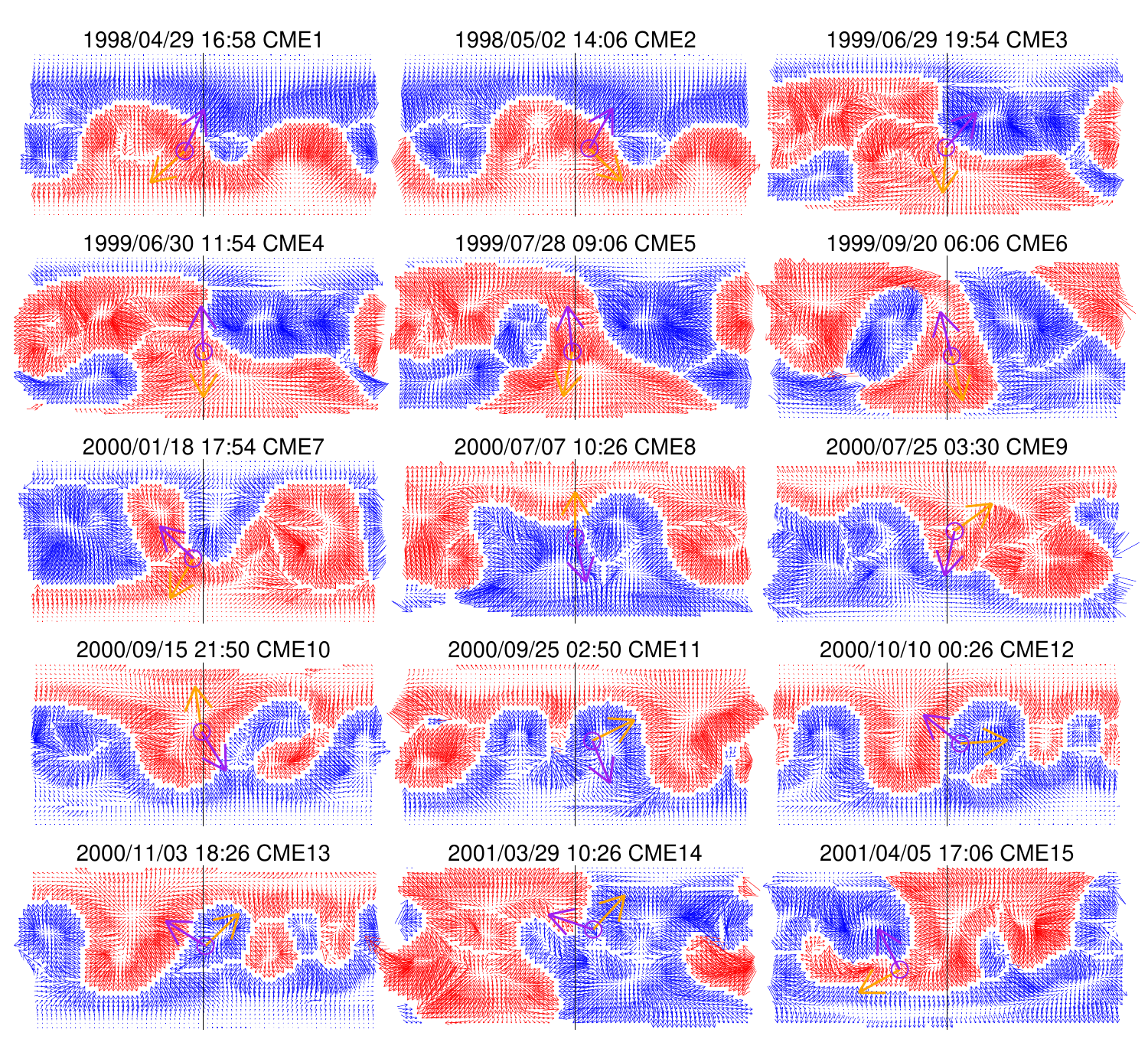}
\caption
{
\small First 15 of the 44 CMEs from the CDAW halo-CMEs list whose deflection angle exceeds 90$^\circ$. The black vertical line in the middle refers to the central meridian at the time of the event. The arrows show the two-dimensional projected vector tracing positive and negative polarity coronal magnetic field (CMF) from 1.1 Rs to the source surface of 2.5 Rs derived using daily synoptic maps of SOHO/MDI and SDO/HMI in blue and red, respectively. The purple circle indicates the source location of each CME. The orange and purple arrows refer to the radial PA and the MPA of the CME that determined using SOHO/LASCO coronagraph observations.
}
\label{case1}
\end{figure}

\begin{figure}[!h]
\centering
\includegraphics[width=1.\textwidth]{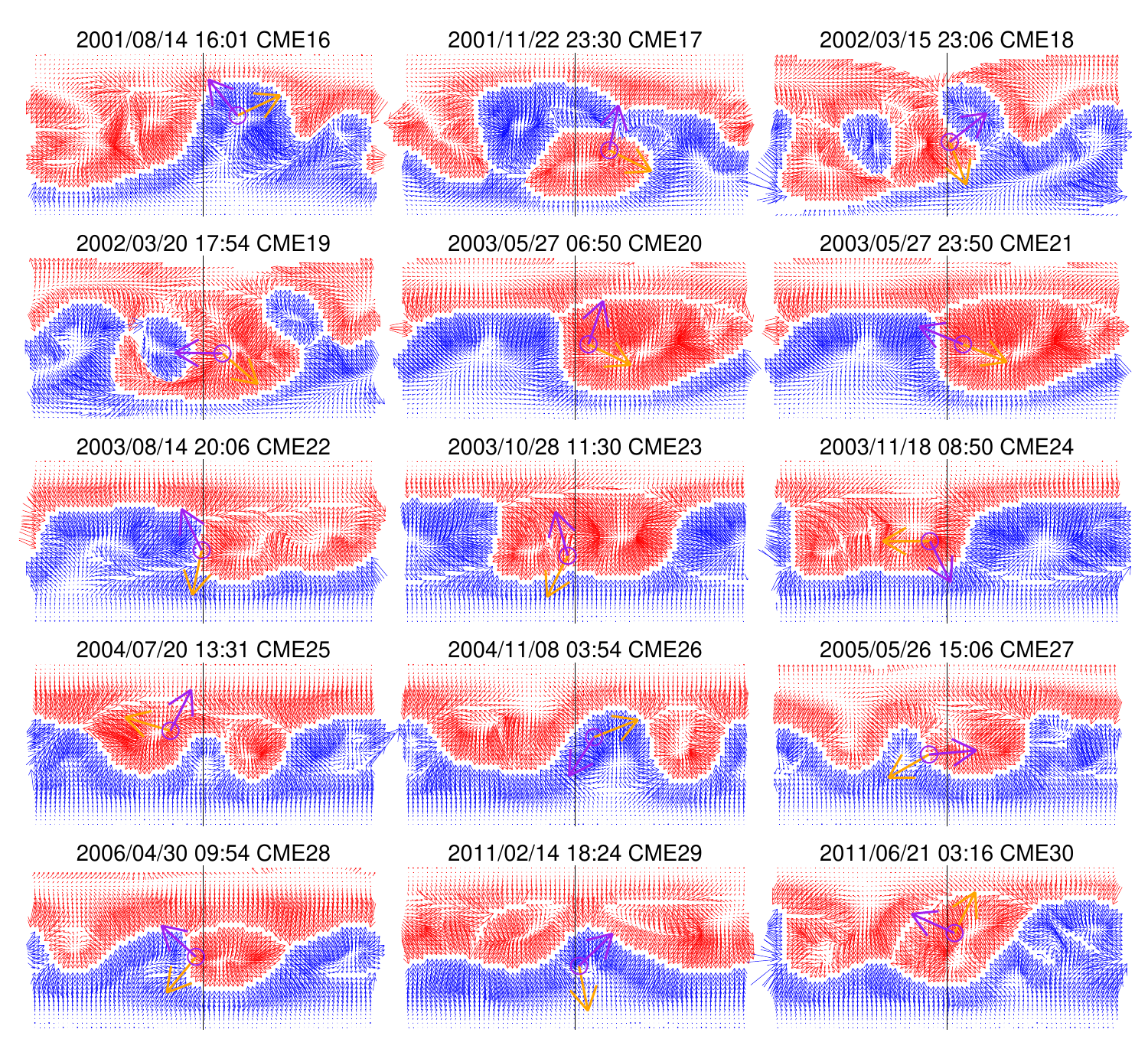}
\caption
{ 
\small CME16 - CME30 from the CDAW halo-CME list whose deflection angle exceeds 90$^\circ$, similar to Figure\, \ref{case1}.
}
\label{case2}
\end{figure}

\begin{figure}[!h]
\centering
\includegraphics[width=1.\textwidth]{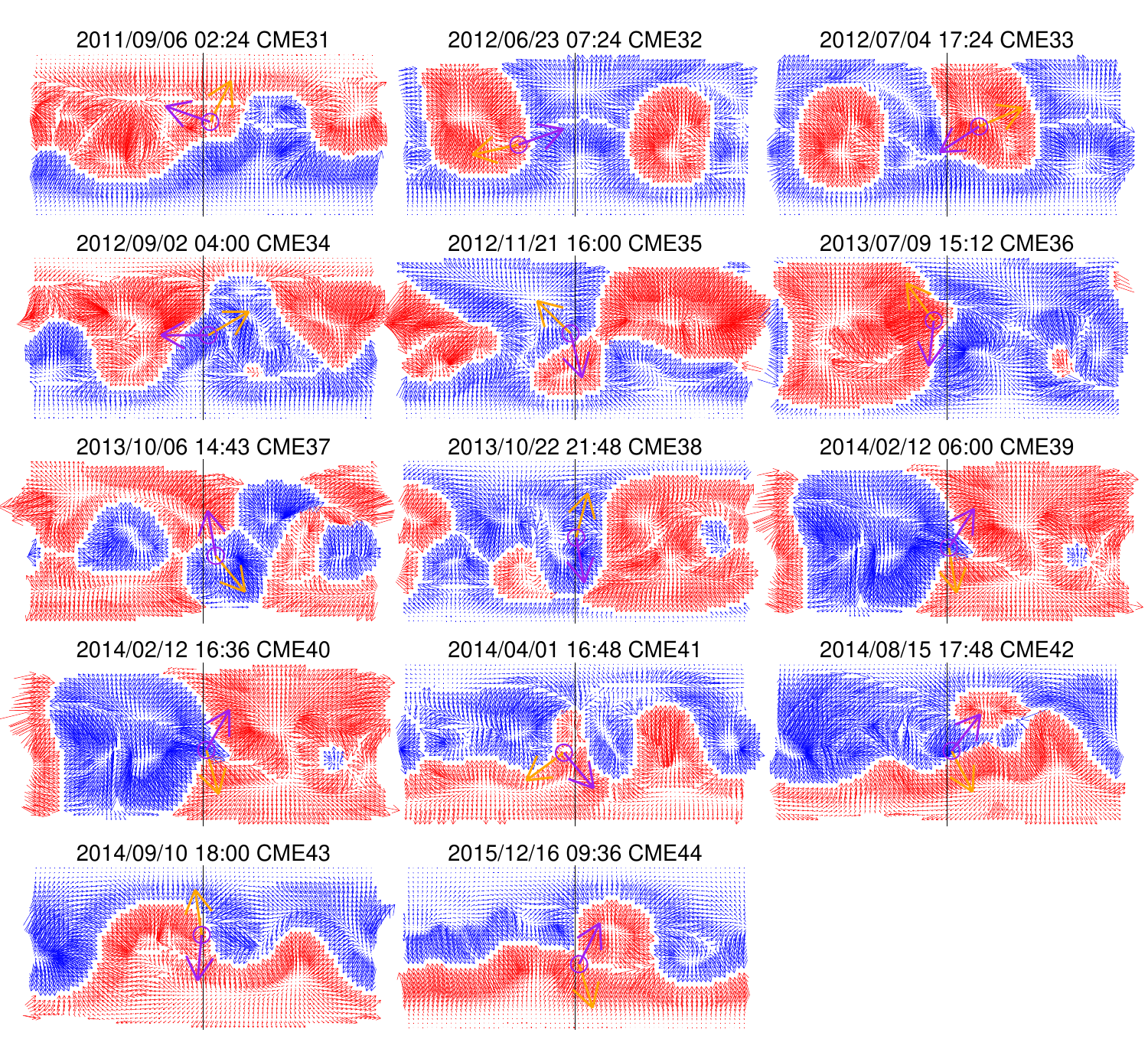}
\caption
{
\small CME31 - CME44 from the CDAW halo-CME list whose deflection angle exceeds 90$^\circ$, similar to Figure\, \ref{case1}.
}
\label{case3}
\end{figure}

If the large-scale coronal magnetic configuration remain the same for several days, it may influence the propagation of multiple CMEs that originate from the same active region, deflecting the CMEs in the same direction. For example, CME1 and CME2 in Figure\, \ref{case1}, which erupted from AR8214 on April 29 and May 2, 1998, are both deflected toward the HCS nearby. In two other cases (CME39 and CME40) in Figure\, \ref{case3}, both CMEs that erupted from AR11974 on Feb 12, 2014 are deflected toward the HCS nearby. However, even if the large-scale magnetic field configuration remains similar, CMEs that originate from the same active region may not be deflected toward the same HCS or PS. For example, in two cases (CME3 and CME4) in Figure\, \ref{case1}, the CMEs that erupted from AR8611 on Jun 29 and Jun 30, 1999, are deflected toward the HCS and PS nearby, respectively. In two other cases (CME20 and CME21) in Figure\, \ref{case2}, the CMEs that erupted from AR10365 on May 27, 2003, are deflected toward the HCS and the PS nearby, respectively. It is still unclear whether a CME will be deflected toward the HCS or the PS nearby when the ambient magnetic field is near a cross point of the HCS and the PS boundaries.

To further analyze the causes of the CME deflection near the Sun, the gradient of the ambient magnetic field lines passing through the CME location should be taken into account for determining the deflection angle in three dimensions.

\acknowledgments 

We thank the SDO/HMI team members who have contributed to the SDO mission for their hard work! We also thank the Coordinated Data Analysis Workshop (CDAW) Data Center for providing the SOHO/LASCO HALO/CME Catalog. JW was supported by National Natural Science Foundation of China (grant 41604149), Beijing Municipal Science and Technology Project (project Z181100002918004).

\bibliographystyle{apj}
\bibliography{CMEdeflectiongt90_v4}


\clearpage

\clearpage

\clearpage

\clearpage

\clearpage
\begin{deluxetable}{lllrrrrrrll}
\tabletypesize{\scriptsize}
\tablecaption{Information table of 44 CMEs from the CDAW halo-CMEs list.
\label{tbl-1}}
\tablewidth{0pt}
\tablehead{
\colhead{} & \multicolumn{5}{c}{Source and projected CME information} & \colhead{Radial} & \multicolumn{2}{c}{Deflection angle} & \colhead{Separation} & \colhead{Toward\tablenotemark{d}}\\
\cline{2-6}
\cline{8-9}
\colhead{No.} & \colhead{Location} & \colhead{Source} &  \colhead{Start time} & \colhead{MPA} & \colhead{$V_{linear}$} & \colhead{PA} & \colhead{minimum\tablenotemark{a}} & \colhead{projected\tablenotemark{b}} & \colhead{angle\tablenotemark{c}} & \colhead{}
}
\startdata
1 & S18E20 & M6.8 flare & 1998.04.29 16:58 & 336 & 1374 & 134 & 27 & -158 & 4 & HCS\\
2 & S15W15 & X1.1 flare & 1998.05.02 14:06 & 331 & 938 & 224 & 21 & 107 & 13 & HCS\\
3 & S14E01 & M1.6 flare & 1999.06.29 19:54 & 320 & 560 & 176 & 14 & 144 & 72 & HCS\\
4 & S15E00 & M1.9 flare & 1999.06.30 11:54 & 3 & 406 & 180 & 15 & -177 & 52 & PS\\
5 & S15E03 & M2.3 flare & 1999.07.28 09:06 & 6 & 462 & 169 & 15 & -163 & 8 & PS\\
6 & S20W05 & filament & 1999.09.20 06:06 & 14 & 604 & 193 & 21 & -179 & 32 & PS\\
7 & S19E11 & M3.9 flare & 2000.01.18 17:54 & 45 & 739 & 151 & 22 & -106 & 74 & PS\\
8 & N04E00 & filament & 2000.07.07 10:26 & 193 & 453 & 0 & 4 & -167 & 164 & \\
9 & N06W08 & M8.0 flare & 2000.07.25 03:30 & 168 & 528 & 307 & 10 & -139 & 9 & HCS\\
10 & N14E02 & C7.4 flare & 2000.09.15 21:50 & 211 & 257 & 8 & 14 & -157 & 30 & HCS\\
11 & N09W18 & M1.8 flare & 2000.09.25 02:50 & 202 & 587 & 297 & 20 & -95 & 44 & PS\\
12 & N01W14 & C6.7 flare & 2000.10.10 00:26 & 51 & 506 & 274 & 14 & 137 & 23 & HCS\\
13 & N02W02 & C3.2 flare & 2000.11.03 18:26 & 57 & 291 & 315 & 3 & 102 & 68 & HCS\\
14 & N20W19 & X1.7 flare & 2001.03.29 10:26 & 71 & 942 & 318 & 27 & 113 & 49 & HCS\\
15 & S24E50 & M5.1 flare & 2001.04.05 17:06 & 27 & 1390 & 120 & 54 & -93 & 18 & HCS\\
16 & N16W36 & C2.3 flare & 2001.08.14 16:01 & 38 & 618 & 296 & 39 & 102 & 11 & HCS\\
17 & S17W36 & M9.9 flare & 2001.11.22 23:30 & 349 & 1437 & 243 & 39 & 106 & 0 & HCS\\
18 & S08W03 & M2.2 flare & 2002.03.15 23:06 & 309 & 957 & 200 & 9 & 109 & 33 & HCS\\
19 & S17W20 & C3.3 flare & 2002.03.20 17:54 & 89 & 603 & 228 & 26 & -139 & 29 & HCS\\
20 & S07W14 & M1.6 flare & 2003.05.27 06:50 & 342 & 509 & 243 & 16 & 99 & 43 & PS\\
21 & S07W17 & X1.3 flare & 2003.05.27 23:50 & 67 & 964 & 247 & 18 & 180 & 20 & HCS\\
22 & S10E02 & C3.8 flare & 2003.08.14 20:06 & 25 & 378 & 169 & 10 & -144 & 59 & \\
23 & S16E08 & X17. flare & 2003.10.28 11:30 & 15 & 2459 & 154 & 18 & -139 & 162 & \\
24 & N00E18 & M3.9 flare & 2003.11.18 08:50 & 206 & 1660 & 90 & 18 & 116 & 3 & HCS\\
25 & N10E35 & M8.6 flare & 2004.07.20 13:31 & 334 & 710 & 73 & 36 & -99 & 55 & PS\\
26 & N08W20 & C7.9 flare & 2004.11.08 03:54 & 148 & 462 & 292 & 21 & -144 & 123 & \\
27 & S11E19 & B7.5 flare & 2005.05.26 15:06 & 275 & 586 & 121 & 22 & 154 & 135 & \\
28 & S10E08 & C1.8 flare & 2006.04.30 09:54 & 47 & 544 & 142 & 13 & -95 & 10 & HCS\\
29 & S20W04 & M2.2 flare & 2011.02.14 18:24 & 315 & 326 & 191 & 20 & 124 & 75 & HCS\\
30 & N16W08 & C7.7 flare & 2011.06.21 03:16 & 65 & 719 & 334 & 18 & 91 & 28 & PS\\
31 & N14W07 & M5.3 flare & 2011.09.06 02:24 & 70 & 782 & 334 & 16 & 96 & 91 & \\
32 & S11E60 & C3.1 flare & 2012.06.23 07:24 & 290 & 1263 & 103 & 61 & -173 & 14 & HCS\\
33 & N14W34 & M1.8 flare & 2012.07.04 17:24 & 124 & 662 & 294 & 36 & -170 & 45 & HCS\\
34 & N03W05 & C2.9 flare & 2012.09.02 04:00 & 90 & 538 & 301 & 6 & 149 & 74 & HCS\\
35 & N05E05 & M3.5 flare & 2012.11.21 16:00 & 194 & 529 & 45 & 7 & 149 & 5 & HCS\\
36 & N19E14 & filament & 2013.07.09 15:12 & 174 & 449 & 35 & 23 & 139 & 123 & \\
37 & S16W13 & C1.1 flare & 2013.10.06 14:43 & 10 & 567 & 218 & 21 & 152 & 6 & HCS\\
38 & N04W01 & M4.2 flare & 2013.10.22 21:48 & 190 & 459 & 346 & 4 & -156 & 71 & PS\\
39 & S12W02 & M3.7 flare & 2014.02.12 06:00 & 328 & 373 & 189 & 12 & 139 & 9 & HCS\\
40 & S11W03 & M2.1 flare & 2014.02.12 16:36 & 331 & 533 & 195 & 11 & 136 & 1 & HCS\\
41 & S09E12 & filament & 2014.04.01 16:48 & 219 & 247 & 127 & 15 & 92 & 158 & \\
42 & S10W05 & filament & 2014.08.15 17:48 & 323 & 342 & 206 & 11 & 117 & 13 & HCS\\
43 & N14E02 & X1.6 flare & 2014.09.10 18:00 & 175 & 1267 & 8 & 14 & 167 & 147 & \\
44 & S13W04 & C6.6 flare & 2015.12.16 09:36 & 334 & 579 & 197 & 14 & 137 & 0 & PS\\
\enddata
\tablenotetext{a}{This column is the minimum deflection angle (angular distance between the CME location and the center of the solar disk) in three dimensions.}
\tablenotetext{b}{This column is the projected deflection angle in two dimensions.}
\tablenotetext{c}{This column is the separation angle between the CME MPA and the direction of the ambient CMF in the plane of the sky.}
\tablenotetext{d}{This column indicates the boundary toward which each CME is deflected. Blank entries are not deflected toward a boundary.}
\label{events}
\end{deluxetable}

\end{document}